\documentclass[conference]{IEEEtran}
\IEEEoverridecommandlockouts
\usepackage{adjustbox}
\usepackage{cite}
\usepackage{amsmath,amssymb,amsfonts}
\usepackage{amsthm}
\usepackage[ruled,vlined,linesnumbered,noresetcount]{algorithm2e}
\usepackage{graphicx}
\usepackage[subtle,paragraphs=normal]{savetrees}
\usepackage{textcomp}
\usepackage{xcolor}
\usepackage{color,soul}
\usepackage{xspace}
\usepackage{booktabs}
\newtheorem{theorem}{Theorem}
\usepackage{optidef}
\usepackage{algpseudocode}
\usepackage{lipsum}
\setstcolor{red}
\usepackage{soul}
\usepackage{tcolorbox}

\def\BibTeX{{\rm B\kern-.05em{\sc i\kern-.025em b}\kern-.08em
    T\kern-.1667em\lower.7ex\hbox{E}\kern-.125emX}}

\begin{document}

\title{Cell Switching in HAPS-Aided Networking: How the Obscurity of Traffic Loads Affects the Decision}   

 

\author{\IEEEauthorblockN{Berk~Çiloğlu\IEEEauthorrefmark{1}$^{,}$\IEEEauthorrefmark{4}, Görkem~Berkay~Koç\IEEEauthorrefmark{1}$^{,}$\IEEEauthorrefmark{4}, Metin~Ozturk\IEEEauthorrefmark{1}$^{,}$\IEEEauthorrefmark{4}, Halim~Yanikomeroglu\IEEEauthorrefmark{4}}
\thanks{{\IEEEauthorrefmark{1}Electrical and Electronics Engineering, Ankara Yıldırım Beyazıt University, Ankara, Türkiye.
\IEEEauthorrefmark{4}Non-Terrestrial Networks (NTN) Lab, Systems and Computer Engineering, Carleton University, Ottawa, Canada.}

This research has been sponsored in part by The Scientific and Technological Research Council of Türkiye (TUBITAK).
}
}


\maketitle

\begin{abstract}
This study aims to introduce the cell load estimation problem of cell switching approaches in cellular networks---specially presented in a high-altitude platform station~(HAPS)-assisted network.
The problem arises from the fact that the traffic loads of sleeping base stations for the next time slot cannot be perfectly known, but they can rather be estimated, and any estimation error could result in divergence from the optimal decision, which subsequently affects the performance of energy efficiency.
The traffic loads of the sleeping base stations for the next time slot are required because the switching decisions are made proactively in the current time slot.
Two different $Q$-learning algorithms are developed; one is full-scale, focusing solely on the performance, while the other one is lightweight and addresses the computational cost.
Results confirm that the estimation error is capable of changing cell switching decisions that yields performance divergence compared to no-error scenarios.
Moreover, the developed $Q$-learning algorithms perform well since an insignificant difference (i.e., 0.3\%) is observed between them and the optimum algorithm.
\end{abstract}


\section{Introduction}
\textit{Ubiquitous connectivity} and \textit{sustainability} are two imperative ingredients of the sixth generation of communication systems~(6G), as they are envisioned to be key aspects in 6G networks by International Telecommunication Union~(ITU)~\cite{itu_vision_june_23}.
To this end, cell switching~(CS), which aims to reduce the energy consumption of base stations~(BSs) by putting them into a sleep mode, has been identified as a viable tool for sustainability in the literature~\cite{sleep1, profit, ELAA2022JR}.
Besides, high-altitude platform station~(HAPS) can be seen as an integral part of CS in cellular networks~\cite{kirmizi_magazin}, as the idle users can be offloaded to HAPS super macro BS~(HAPS-SMBS)~\cite{smbs}; providing an extra capacity for more switching off opportunities and serving as a new BS where existing infrastructure is not sufficient.
The integration of non-terrestrial networks (NTNs), through HAPS, with terrestrial networks (TNs), therefore, pave the way for ubiquitous connectivity and sustainability in 6G networks.

The literature is not ripe in terms of HAPS-assisted CS in cellular networks with only a few exceptions.
For example, the study in~\cite{bizim}, which is the earlier version of this current paper, proposed CS for HAPS-integrated TN with a simple exhaustive search~(ES) algorithm.
The authors in~\cite{yeni_haps} investigated the case where the entire load was offloaded to the HAPS-SMBS instead of the macro cell, while in~\cite{meryem} sorting of small BS~(SBS) payloads in ascending order was studied to transfer traffic to macro BS and HAPS-SMBS. 
On the other hand, the CS literature on TNs is quite saturated with a plethora of distinctive studies.
The authors in~\cite{ELAA2022JR} studied various sleep depth levels, while the work in~\cite{metin} applied the SARSA algorithm with value function approximation to reduce the energy consumption. 

Although the literature has been full of CS methods taking the problem from various angles, an overlooked fact exists regardless of being TN-only or TN-NTN integration: cell load estimation problem.
To the best of our knowledge, this problem has never been identified in the literature before; however, if not addressed, it has the potential of making many studies inapplicable---setting a barrier between cellular communication networks and sustainability goals.
The cell load estimation problem originates from the fact that CS studies in the literature are predominantly dependent on the cell loads~\cite{ELAA2022JR,bizim,meryem,yeni_haps,metin}; i.e., the CS decision/optimization is performed using the traffic loads of the cells.
Nevertheless, the problem is that once a BS is put on a sleep mode at a certain time slot, how can we know its traffic load at the preceding time slot in order to decide its ON/OFF status?
Further, in a more deeper understanding, how can we know the traffic loads of the cells within a time slot while performing a search-based optimization?
The fact is that the loads cannot be perfectly known but can be estimated with accompanying errors.

This work is the very first attempt to introduce the cell load estimation problem in the CS literature in cellular networks.
Moreover, after introducing this novel problem, we build a CS optimization model by considering a HAPS-SMBS in the network (i.e., TN-NTN integration), followed by developing two different $Q$-learning algorithms to solve the optimization problem. 
We then investigate the performance of the developed algorithms with and without the estimation errors in order to observe and reveal the impacts of the cell load estimation problem on the CS performance.
In this regard, the contributions of this study are given as follows:
\begin{itemize}
    \item The cell load estimation problem is introduced for the first time in the literature with its mathematical foundations.
    \item We consider a HAPS-SMBS in the network for vertical offloads in an effort to provide \textit{TN-NTN integration} and \textit{sustainability} as envisioned in 6G networks.
    \item Full-scale and lightweight $Q$-learning algorithms are designed for accuracy and computational cost purposes, respectively.
    \item We investigate the results under cell load estimation errors to understand their effects on the CS performance.
\end{itemize}

\subsection{Notation}
The following notation is adopted throughout the paper: indices/subscripts $i$, $j$, and $k$ denote users/UEs, BSs (HAPS-SMBS and SBSs), and SBSs, respectively.
\textbf{Bold} characters in mathematical equations indicate a vector or matrix.
Subscripts that are variables are used in \textit{italic} form while those that are non-variables and used for naming purposes only are given in straight form.
Superscripts are used to differentiate the cases/scenarios for the same variable.
\section{System Model}
\subsection{Network Model}
\begin{figure}
    \centering
    \includegraphics[width=.8\linewidth,trim={6cm 1cm 4cm 1cm},clip]{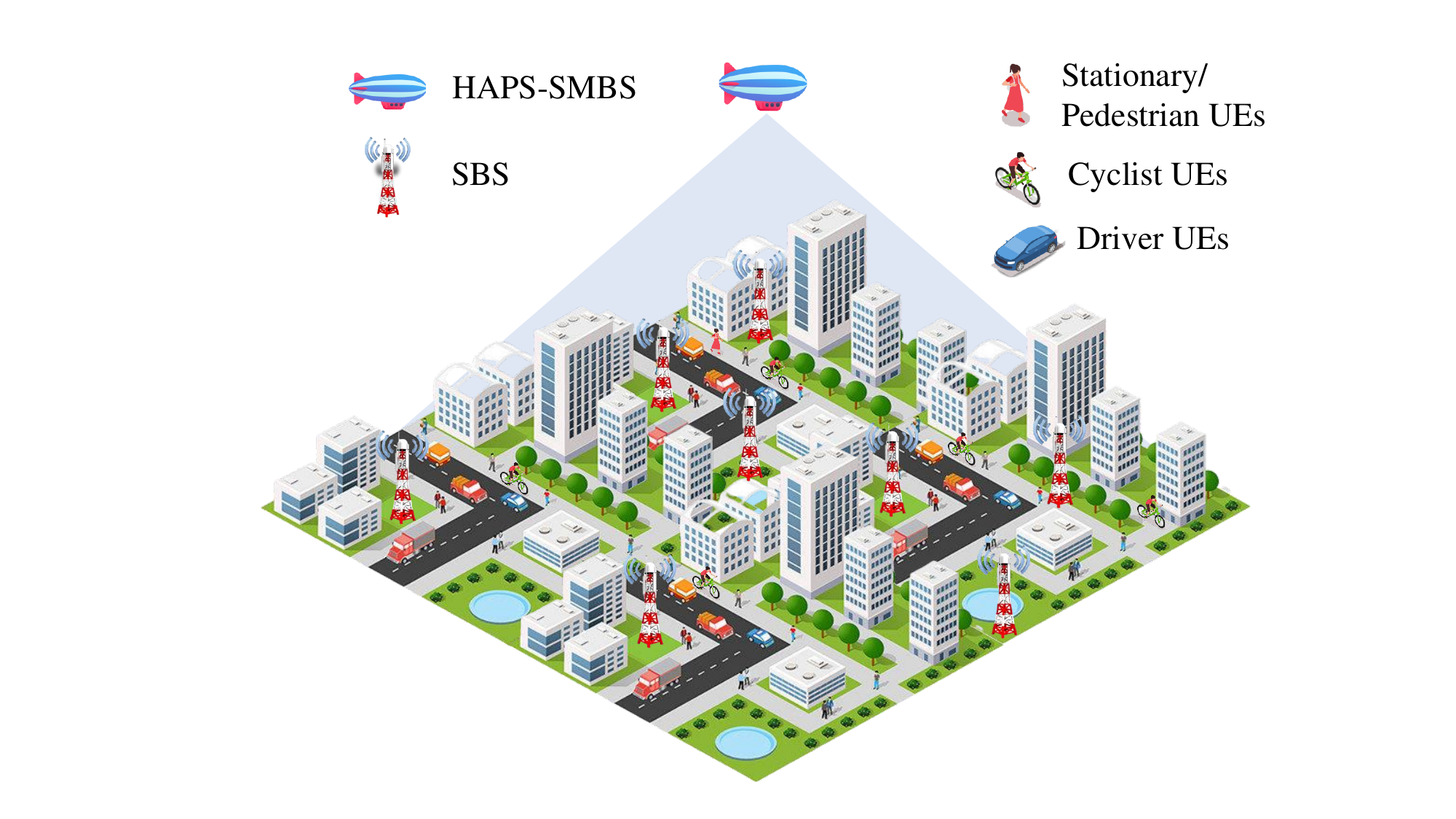}
    \caption{A typical VHetNet model containing SBSs, HAPS-SMBS, and UEs.}
    \label{fig:networkmodel}
\end{figure}
A vertical heterogeneous network~(VHetNet) including $n \in \mathbb{N}$ SBSs, with $k=1,~2,~...,~n$ being the indices (i.e., IDs) of SBSs, and also a single HAPS-SMBS is considered as a network model. 
There are totally $n+1$ BSs, with $j =1,~2~,~...,~(n+1)$ being the indeces of BSs, are deployed consisting of $n$ SBSs and a single HAPS-SMBS. 
There are also $e \in \mathbb{N}$ users/UEs, with $i=1,~2,~...,~e$ being the indices for the users/UEs, distributed around the network environment.
We also introduce a user density variable, $\delta$, which is defined by the number of users per m$^2$, such that $\delta=e/A$ where $A$ is the area (in m$^2$) of the network.
SBSs are placed in a symmetrical manner with reference to the center of the environment and HAPS is deployed from 20 km above the ground by centralizing the environment.
Due to the HAPS-SMBS's enormous footprint, it is not realistic to assume that the entire capacity of HAPS-SMBS is dedicated to our considered network, thereby we set the utilized capacity of HAPS-SMBS to a certain level: 70\% of the HAPS-SMBS capacity is reserved for the network-in-question, while the rest of 30\% is utilized by other networks within the footprint of HAPS-SMBS.
The system model is illustrated in Fig.~\ref{fig:networkmodel}, wherein different UE mobility modes are considered: stationary, pedestrian, cyclist, and driver.
\subsection{Propagation Model}
The path loss model of HAPS-SMBS is built as described in the 3GPP report given in~\cite{3GPP_HAPS}.
Since shadowing could affect path loss in various ways, it is necessary to consider the line-of-sight (LoS) and non-LoS (NLoS) situations.
The path loss, $L^\omega$, $\omega \in \text{\{LoS,~NLoS\}}$, is calculated as~\cite{3GPP_HAPS}
\begin{equation}\label{eq:PL2}
    L^\omega = L^\omega_\text{b} + L_\text{g} + L_\text{s} + L_\text{e},
\end{equation}
where $L^\omega_\text{b}$ is the basic path loss, $L_\text{g}$ is the attenuation caused by atmospheric gasses, $L_\text{s}$ is the loss incurred by either ionospheric or tropospheric scintillation, and $L_\text{e}$ represents building entry loss. 
More details about the path loss model of HAPS-SMBS are given in~\cite{bizim}.
For the consistency between NTN and TN path loss models, the 3GPP report given in~\cite{3GPP_Terrestrial} is employed for TN.

\subsection{User-Cell Association}\label{sec:user_assoc}
The user-cell association is conducted by considering the signal-to-interference-plus-noise ratio (SINR) levels, such that to associate user $i$ to BS $j$, there are three certain criteria to be satisfied.
Let these criteria be $\kappa_1$, $\kappa_2$, and $\kappa_3$, where $\kappa_1,~\kappa_2,~\kappa_3 \in \{0,1\}$ in such a way that if they are satisfied they get the value of 1 and 0 otherwise.
First, a BS needs to provide the highest SINR to the UE; $\kappa_1=\{\varkappa_1\in \{0,1\}=1~|~\gamma_{j} > \forall \gamma_{h\neq j},~(1 \leq j,~h \in \mathbb{N}\leq n+1)\}$, where $\gamma_j$ denotes the instantaneous SINR of BS $j$ at time~$t$.
Second, the BS needs to have an available capacity to accommodate UE $i$: $\kappa_2=\{\varkappa_2\in \{0,1\}=1~|~\Omega^\text{a}_j \geq D_i\}$, where $\Omega^\text{a}_j\in \mathbb{R^+}$ and $D_i\in \mathbb{R^+}$ represent the available capacity of BS $j$ and the demand of UE $i$ at time $t$, respectively, while $\Omega_j\in \mathbb{R^+}$ indicates the total capacity of BS $j$.
Third, the receiver sensitivities of UEs have to be respected to assure the quality-of-service (QoS); i.e., $\kappa_3=\{\varkappa_3\in \{0,1\}=1~|~P_{\text{RX},i} \geq P_{\text{RX}_\text{min},i}\}$, where $P_{\text{RX},i}$ and $P_{\text{RX}_\text{min},i}$ denote the received power and receiver sensitivity of UE $i$ at time $t$, respectively.

Let $U_{i,j}\in \{0,1\}$ denote the association between UE $i$ and BS $j$, such that $U_{i,j}=1$ if UE $i$ is associated to BS $j$, and $U_{i,j}=0$ otherwise.
Thus, the user-cell association is given by 
\begin{equation}\label{eq:assoc}
U_{i,j} = 
    \begin{cases}
        1, & (\kappa_1=1)\land(\kappa_2=1)\land(\kappa_3=1),\\
        0, & \text{otherwise}.
    \end{cases}
\end{equation}
\subsection{Data Traffic Model}
Radio resources are modeled based on resource blocks~(RBs); i.e., the total capacities of BSs (SBSs or HAPS-SMBS) are divided into RBs.
At each time slot, each UE generates data traffic with a certain rate, which requires a certain amount of bandwidth (i.e., RBs).
Put simply, each UE demands a certain amount of RBs with a random process: $r_i = \Psi\times \mathcal{U}\{r_\text{l},~r_\text{u}\}$, where $\Psi$ is a coefficient to shape the RB demand profile of UEs.
$r_\text{l}$ and $r_\text{u}$ are the lower and upper bounds that a UE can demand at time $t$, and here they are selected as 1 and 3, respectively.
Note that $\Psi$ is obtained from a real data set provided by Telecom Italia for Milan, Italy~\cite{milan}.

\subsection{Power Consumption Model}
The power consumption model is adopted from the EARTH model in~\cite{earth} to calculate $P_\text{B}^\alpha$ where $\alpha \in \text{\{S,~H\}}$.
Hereafter, any variable with superscript S and H indicates that the variable is for SBSs and HAPS-SMBS, respectively. 
The power consumption of a BS is given by
\begin{equation}\label{eq:pow1}
P_\text{B}^\alpha = 
    \begin{cases}
        P_\text{C}^\alpha + \xi^\alpha \rho P_\text{max}^\alpha, & 0 < P_\text{TX}^\alpha < P_\text{max}^\alpha, \\
        P_\text{S}^\alpha, & P^\alpha_\text{TX}= 0,
    \end{cases}
\end{equation}
where $P_\text{C}^\alpha$ is constant power, $\xi^\alpha$ slope of the load-dependent power consumption value, $\rho=[0,1]$ is the load factor of BSs.
$P_\text{TX}^\alpha$ and $P_\text{max}^\alpha$ are the instantaneous and the maximum transmit powers of BSs, respectively.
$P_\text{S}^\alpha$ is the power consumption when a BS is put into the sleep mode. 

The total load factor of the network, $\rho_\text{T}$, can be defined as $\rho_\text{T}=\sum_{j=1}^{n+1}\rho_j$.
Here, $\rho_j$ can be expressed as $\rho_j=\sum_{i=1}^{e_j}r_i$,
where $e_j$ is the number of UEs that BS $j$ accommodates. 
Therefore, $\rho_\text{T}$ can be rearranged as
\begin{equation}
    \rho_\text{T}=
        \sum_{j=1}^{n+1}\sum_{i=1}^{e_i}r_i,
\end{equation}
making $\rho_\text{T}$ a function of the number of BSs, $n+1$, the user density, $\delta$, and the RB demand of UEs, i.e., $\rho_\text{T}=f(n,~\delta,~r)$, as the user density is a function of the number of users, such that $\delta=f(e)$.
It is important to note here is that $e = \sum_{j=1}^{n+1}e_j~\text{iff}~\sum_{j=1}^{n+1}\Omega_j \geq \sum_{i=1}^{e}r_i$, yielding the possibility of unconnected/uncovered users.

\section{Problem Formulation}\label{sec:problem}
This section describes the problem in a formal way by providing an optimization modeling and mathematical derivations.

\begin{theorem}\label{th:est}
    The error caused by the traffic load estimation, $\varepsilon \in \mathbb R$, can change the policy, $\pmb \eta$, in the optimization process, which is based on the cell loads; i.e., the objective function of the optimization problem is a function of cell loads.
\end{theorem}
\begin{proof}
First, it is important to give the optimization model in order to reveal what kind of problem we work on. 
This work primarily focuses on minimizing the total power consumption of a VHetNet, consisting of several SBSs and a single HAPS-SMBS, by finding the best switching off policy.
A policy $\pmb \eta_t = [\beta_{1,t},~\beta_{2,t},~\beta_{3,t},~...,~\beta_{n,t}]$ indicates which SBSs are ON or OFF at a given time $t$, where $\beta_{k,t} \in \{0,1\}$  denotes the ON/OFF status of SBS $k$ at time $t$. 
HAPS-SMBS is assumed to be always active.

The combinatorial problem for the considered scenario, with multiple SBSs and a single HAPS-SMBS, can be modeled as~\cite{ELAA2022JR}
\begin{mini}|s|
    {\eta}{P_\text{N}} 
    {\label{eq:opt}}{}
    \addConstraint{\rho}{\leq1}{\qquad\text{(C$_1$)}}
    \addConstraint{P_\text{TX}^\alpha}{\leq P_\text{max}^\alpha}{ \qquad\text{(C$_2$)},}
\end{mini}
where C$_1$ restricts the capacities of both SBSs and HAPS-SMBS from being violated while C$_2$ prevents BSs (SBS or HAPS-SMBS) from exceeding their maximum transmit power.
$P_\text{N}$ is the total power consumption of the considered VHetNet, wherein several SBSs and a single HAPS-SMBS co-exist, and can be expressed as
\begin{equation}\label{eq:totalpower}
P_\text{N} = P^\text{H}_\text{B} + \sum_{k=1}^{n}P^\text{S}_{\text{B},k},
\end{equation}
where $P^\text{S}_{\text{B},k}$ indicates the power consumption of the $k^\text{th}$ SBS.

Substituting \eqref{eq:pow1} into \eqref{eq:totalpower} yields

\begin{equation}\label{eq:closed-form-power}
    P_\text{N} =\Big[P^\text{H}_\text{B}+\sum_{k=1}^{n} (P^\text{S}_{\text{C},k} + \xi^\text{S}_k \rho_k P^\text{S}_{\text{max},k})\beta_{k,t} + P^\text{S}_{\text{S},k}(1-\beta_{k,t}) \Big].
\end{equation}
Since the optimal policy, $\pmb \eta_\text{opt}$, is the one that minimizes $P_\text{N}$ by satisfying C$_1$ and C$_2$, such that
\begin{equation}\label{eq:arg-min}
      \arg \min_{\pmb\eta}(P_\text{N})=\pmb\eta_\text{opt},
\end{equation}
the decision (i.e., the policy, $\pmb \eta$) is taken based on the total power consumption of the network, $P_\text{N}$, and thereby on the load factor, $\rho$.
In other words, from~\eqref{eq:closed-form-power}, $P_\text{N}=f(\rho)$ and, from~\eqref{eq:arg-min}, $\pmb\eta_\text{opt}=g(P_\text{N})$, where $f$ and $g$ are functions.
Then we can express $\pmb\eta_\text{opt}$ as a function of $\rho$ by composing $f$ and $g$, such that $\pmb\eta_\text{opt}=g(P_\text{N})=g(f(\rho))$ where $(g \circ f) (x)=g(f(x))$.

Therefore, an error, $\varepsilon$, in the load factor would change the decision/policy; i.e., if $\varepsilon$ is added to $\rho$ in~\eqref{eq:closed-form-power}, we get an erroneous (i.e., updated with error) power consumption as
\begin{equation}\label{eq:closed-form-power-updated}
    P'_\text{N} =\Big[P^\text{H}_\text{B}+\sum_{k=1}^{n} (P^\text{S}_{\text{C},k} + \xi^\text{S}_k (\rho_k+\varepsilon) P^\text{S}_{\text{max},k})\beta_{k,t} + P^\text{S}_{\text{S},k}(1-\beta_{k,t}) \Big],
\end{equation}
which in turn changes the value in~\eqref{eq:arg-min} (i.e., updates the optimal policy: $\pmb\eta'_\text{opt}$) as $\arg \min_{\pmb\eta}(P'_\text{N})=\pmb\eta'_\text{opt}$, where $\exists\pmb\eta'_\text{opt}=\pmb\eta_\text{opt}$.
Therefore, when the CS optimization model includes cell loads in the decision process, the errors that occur while estimating the loads of the sleeping SBSs for the next time slot make the process erroneous, which has a potential of altering the final decision.
\end{proof}

\subsection{Cell Load Estimation Problem}
The optimization problem given in~\eqref{eq:opt} ensures that the total power consumption is minimized by respecting the given constraints.
Since this is a combinatorial optimization problem, there are switching possibilities, in which there exists a policy (i.e., the best policy, $\pmb\eta_\text{opt}$) that minimizes the total power consumption, $P_\text{N}$.
On one hand, the optimization objective in~\eqref{eq:opt} yields that the decision of selecting the best policy, $\pmb\eta_\text{opt}$, actually depends on the load factors of BSs, $\rho$.
On the other hand, since this is a combinatorial optimization problem, search or metaheuristic algorithms are employed for the solution, meaning that they have to try different combinations of active and deactive SBSs and find the best one.
From there, it is worth noting that the problem is two-fold: inter-time-slot and intra-time-slot.
\subsubsection{Inter-Time-Slot Load Estimation Problem}
The question is how can we know the traffic load of an already sleeping SBS for the next time slot?
In other words, determining the load factor information, $\rho_k$, of SBS $k$, which was previously scheduled to enter sleep mode in the previous time slot, is essential for deciding its status in the next time slot. However, existing literature on CS predominantly assumes perfect knowledge about traffic loads~\cite{ELAA2022JR,meryem,yeni_haps,bizim,metin}.
The traffic loads of sleeping cells for the next time slots are needed due to the fact that CS decisions are made for one slot ahead, meaning that the decision for the next time slot, $t+1$, is made at the current time slot, $t$.
Otherwise, at the beginning of each time slot, all the SBSs need to be switched on to get their traffic information, $\rho$, followed by the CS optimization, which makes the process inefficient and hard to implement.
\subsubsection{Intra-Time-Slot Load Estimation Problem}
Search and heuristic algorithms, such as ES, genetic algorithm, etc., need to search different switching options to find the best one (i.e., the least power consuming one).
Nonetheless, how come the traffic loads of different switching options are acquired?
The most accurate approach would be performing each option in real time, but this is quite infeasible, because when the number of SBSs is large, the search process would take long so that the switching decision cannot be taken within the time slot frame. Additionally, performing each combination in real time means that user-cell association is conducted repeatedly even within a time slot, resulting in repeated connection switching that subsequently dissatisfies the user.

In short, whether it be inter-time-slot or intra-time-slot, the fundamental fact is that the traffic loads of SBSs cannot be known perfectly; such assumptions are impractical and unrealistic.
The traffic loads should rather be estimated, which yields estimation errors, $\varepsilon$, in return.
To this end, \textit{the hypothesis behind this work is that estimation errors are capable of changing the decision (i.e., the switching off policy, $\pmb \eta$) in the optimization process}.

\section{Methodology}
In addition to utilizing the ES algorithm as a baseline for the CS operations, we propose to employ $Q$-learning algorithm, which is based on the consequences of actions, in other words, penalties or rewards. 
$Q$-learning uses a matrix (i.e., $Q$-table) to log the results of all the action-state pairs, making it prone to create problems when the number of actions and/or states goes high.
Considering this, in this work, we propose two different $Q$-learning designs: full-scale and lightweight.
The former is more accurate as the states are made equal to all the combinations of ON/OFF status of SBSs, while the latter is simplified with redefined states---significantly reducing the number of states and, subsequently, the computational cost.

A central controller unit (CCU), located at the HAPS, comprises a processor and a database, serving various functions including acquiring cell loads from active SBSs, estimating cell loads for sleeping SBSs, and implementing $Q$-learning algorithms.
Essentially, the CCU acts as an agent for $Q$-learning. 
When an SBS is active, it shares its traffic with the CCU via a control channel, and the traffic loads for all SBSs in the network are stored in the CCU's database. 
Consequently, when an SBS enters sleep mode, its traffic load can be estimated by the processor using various methodologies, such as statistical or machine learning approaches, leveraging historical data stored in the database.

\subsection{Proposed $Q$-learning Designs}\label{sec:proposed_q}
\begin{figure} [h!]
   \centering
   \includegraphics[width=.55\linewidth,trim={6cm 3cm 8cm 2cm},clip]
   {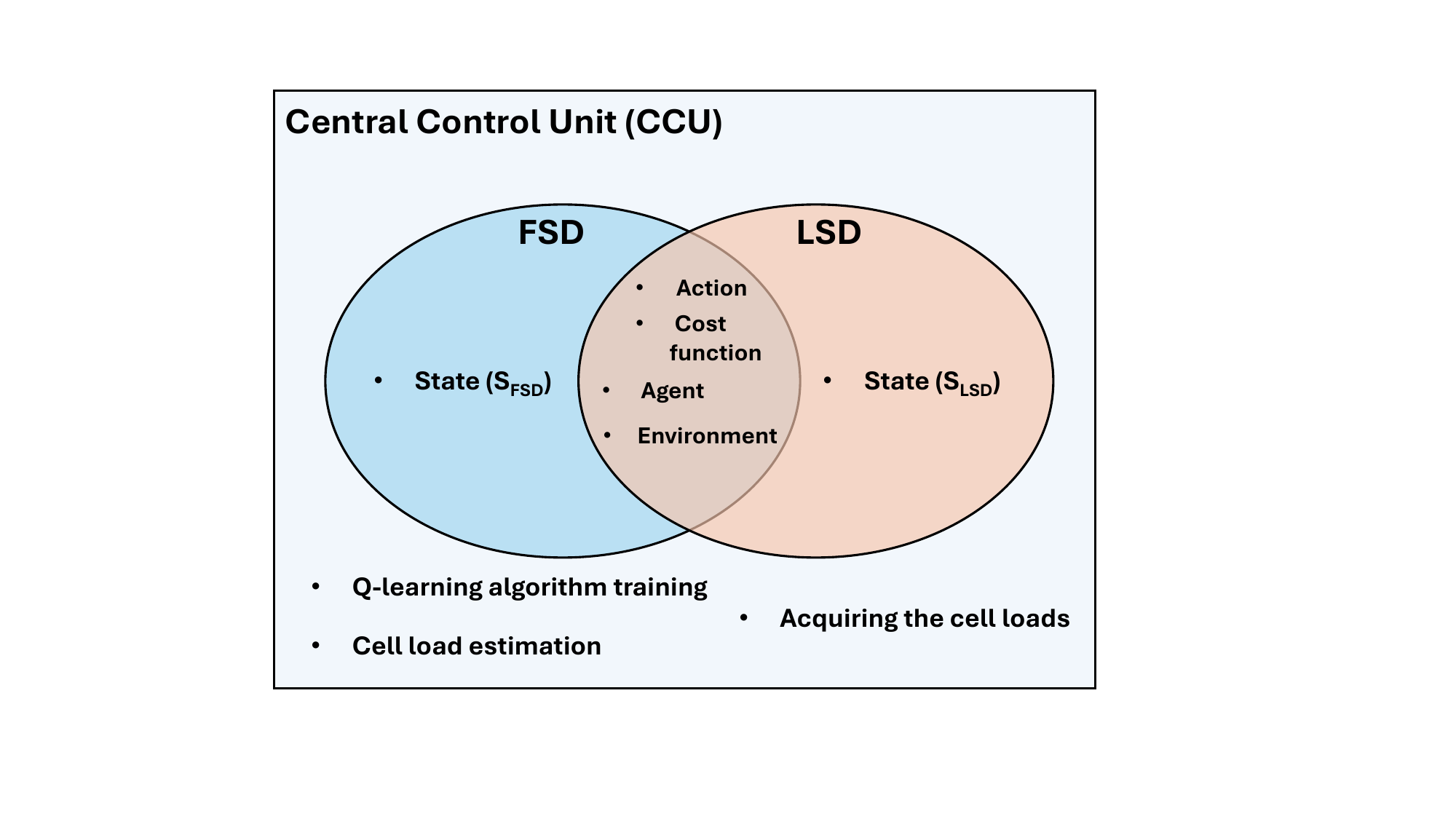}
    \caption{The architectural commonalities and differences between the designs.}
    \label{fig:networkmodel}
    \label{fig:venn}
\end{figure}

We consider two different $Q$-learning designs; namely full-scale state design (FSD) and lightweight state design (LSD).
The relationship between the algorithms can be visualized as in Fig.~\ref{fig:venn}.

\subsubsection{Cost Function}
The cost function, which is common for both $Q$-learning designs, is given by
\begin{equation}\label{eq:cost}
    \Gamma(\zeta,~P_\text{N}) =  \zeta \times P_\text{N} + (1-\zeta)\times \Phi,
\end{equation}
where $\zeta=0$ if $\sum_i \sum_j U_{i,j}<e$ and $\zeta=1$ if $\sum_i \sum_j U_{i,j}=e$.
Here, $\Phi$ is an astronomical constant penalty to avoid choosing an action that results in an unconnected UE, i.e., $\sum_i \sum_j U_{i,j}<e$.

\subsubsection{Full-Scale State Design (FSD)}
In this design, the actions, $\pmb A_\text{FSD}$, and the states, $\pmb S_\text{FSD}$, are equal (with dimensions of $[2^n\times n]$) and composed of all possible SBS activation/deactivation combinations.
The state matrix, $\pmb S_\text{FSD}$, is given by
\begin{equation}\label{eq:state_fsd}
   \pmb S_\text{FSD} =   \begin{pmatrix}
 0 & 0 & ...& 0&0 \\
 0 & 0 & ...& 0&1\\
& &. \\
& &.  \\
& &.  \\
1&1&...&1&1 \\
\end{pmatrix},
\end{equation}  
which is also equal to $\pmb A_\text{FSD}$. The cost function is as given in~\eqref{eq:cost}, and the computational complexity for this design becomes $O(2^{2n})$.

\subsubsection{Lightweight State Design (LSD)}
This design is used to simplify the $Q$-table in FSD since it is not capable of dealing with large matrix sizes. 
Compared to FSD, only the set of states, $\pmb S_\text{LSD}$, changes while the actions remain the same, such that $\pmb A_\text{LSD}=\pmb A_\text{FSD}$.
The state definition becomes the number of BSs exceeding the full capacity; i.e., $s_m=\sum_{j=1}^{n+1}\Lambda_j$, where $m=1,~2,~...,~(n+1)$ and $\Lambda_j\in \{0,1\}$ is defined as
\begin{equation}
     \Lambda_j=\begin{cases}
        1,& \sum_{i=1}^{e_j}r_i>\Omega_j,\\
        0,& \sum_{i=1}^{e_j}r_i\leq \Omega_j.
    \end{cases}
\end{equation}
Here, $s_m$, a component of the row-vector $\pmb S_\text{LSD}$ that has a dimension of $[1\times (n+1)]$, indicates the instantaneous state when the action $\pmb a_{(1\times n)}$ is taken, i.e.,
\begin{equation}\label{eq:state_fsd}
   \pmb S_\text{LSD} =  \begin{pmatrix}
s_1 \\
s_2 \\
...\\
s_{(n+2)}
\end{pmatrix}.
\end{equation} 
The cost function is as given in~\eqref{eq:cost}, the computational complexity of this design is $O(2^{n})$.

\subsubsection{Training Process}
In the training process, the agent (i.e., the CCU) takes an action from $\pmb A_\text{FSD}$ or $\pmb A_\text{LSD}$, followed by calculating the cost, $\Gamma$, and updating the $Q$-table by $Q(s_t, \pmb a_t) \leftarrow Q(s_t, \pmb a_t) + \alpha \left[ \Gamma_{t+1} + \chi \min_a Q(s_{t+1}, \pmb A) - Q(s_t, a_t) \right]$ where subscript $t$ and $t+1$ are used to indicate \textit{current} and \textit{next}, respectively, and $\chi$ is the discount factor.
The actions are taken based on an $\epsilon$-greedy policy, where the agent sometimes takes random actions (following the uniform distribution) with a probability of $\epsilon=[0,1]$ (referred to as \textit{exploration}), and sometimes more informed actions, i.e., the best one in the $Q$-table, (referred to as \textit{exploitation}).
Additionally, the $\epsilon$-decaying approach is adopted, where the value of $\epsilon$ is reduced at every time slot by a certain amount in a way that $\epsilon \leftarrow \epsilon \tau$ is implemented at each time slot, where $\tau$ is the $\epsilon$-decaying factor.
\section{Performance Evaluation}

To evaluate the impacts of load estimation problem, we use MATLAB and run two different simulation campaigns for both small- and large-scale network perspectives.
In the former, $n=4$ SBSs are utilized, while, in the latter, $n=8$ SBSs are considered.
The ES algorithm and all-active approach (A3), where all SBSs are always kept ON, are used as benchmarks.
ES, FSD, and LSD algorithms are accompanied by subscripts ``${\varepsilon=0}$" or ``${\varepsilon>0}$" according to the existence of estimation error, such that if an algorithm is run with the estimation error, then it gets the subscript of ``${\varepsilon>0}$", and vice versa.
Both communication- and $Q$-learning-related simulation parameters are given in Table~\ref{table:simparameter}.

\begin{table}[ht]
\centering
\caption{Simulation Parameters} \label{table:simparameter}
\begin{adjustbox}{width=.85\columnwidth,center}
\begin{tabular}{ll}
\textbf{Parameters} & \textbf{Values} \\ \hline
Environment area ($A$)    & 1025 m $\times$ 1025 m     \\
Time slot number ($N_\text{TS}$)   & 50 \\            
Time slot duration ($t_\text{d}$)  & 1 s             \\
$\varepsilon$ ranges & $\varepsilon_1$ = [20,40], $\varepsilon_2$ = [60,80], $\varepsilon_3$ = [180,200] \\
$\delta$ for 4SBs & \{100,200,300\} \\
$\delta$ for 8SBs & \{200,400,600\} \\
Carrier frequency ($f_\text{c}$)   & 2.5 GHz         \\
Bandwidth  ($W$)         & 50 MHz          \\
Bandwidth per UE    & 200 kHz         \\
SBSs transmit power    & 33 dBm        \\
HAPS-SMBS transmit power   & 49 dBm   \cite{3GPP_Terrestrial}        \\
SBSs antenna gain    & 4 dBi          \\
HAPS-SMBS antenna gain  & 43.2 dBi   \cite{3GPP_HAPS}      \\
UE antenna gain     & 0 dBi  \cite{3GPP_Terrestrial}          \\
$\sigma^\text{LoS}$ \& $\sigma^\text{NLoS}$ &  4 dB \& 6 dB \cite{3GPP_Terrestrial}  \\
Receiver reference sensitivity  &  -95 dBm  \\
SBSs constant power ($P_\text{C}^\text{SBS}$)             & 56 W           \\
HAPS-SMBS constant power ($P_\text{C}^\text{H}$)            & 130 W        \\
SBSs slope of load-dependent power ($\xi^\text{SBS}$)             & 2.6               \\ 
HAPS-SMBS slope of load-dependent power ($\xi^\text{H}$)             & 4.7                \\ 
SBSs maximum transmit power ($P_\text{max}^\text{SBS}$)    & 6.3 W \\
HAPS-SMBS maximum transmit power ($P_\text{max}^\text{H}$)    & 20 W \\
Sleep mode power ($P_\text{s}$)    & 39 W \\
Astronomical constant penalty ($\phi$) & $10^9$ \\
Learning rate    & 0.9\\
Discount factor ($\chi$) (FSD \& LSD)  & 0.9  \& 0.3\\
Initial $\epsilon$ value & 0.8\\
$\epsilon$-decaying factor ($\tau$) & 0.9 \\
Number of iterations & 20000\\
\bottomrule
\end{tabular}
\end{adjustbox}
\end{table}
\textit{Energy consumption} and \textit{relative difference} are taken as metrics to evaluate the performance of all the considered algorithms and the impacts of load estimation errors.
The energy consumption values are obtained through~\eqref{eq:pow1}, while
relative difference is defined as the percentage differences between the energy consumption values of any two CS algorithms, where the difference between the greater (always mentioned first) and lower values is divided by the greater one.
The network is assessed for different UE densities, $\delta$, which are varied in accordance with the number of UEs.
In this regard, the number of UEs is chosen as 100, 200, and 300 for $n=4$, which are duplicated when $n=8$. 
Moreover, three different $\varepsilon$ ranges, consisting of $\varepsilon_1=[20\%,~40\%]$, $\varepsilon_2=[60\%,~80\%]$, and $\varepsilon_3=[180\%,~200\%]$, are considered. 
These ranges of $\varepsilon$ are the percentages of the actual load factor of a BS $j$, $\rho_j$; in the $\varepsilon_1$ regime, for example, the value of $\varepsilon$ is determined randomly between $20\%$ and $40\%$ with a uniform distribution, followed by multiplying by $\rho_j$. 
To be more generic, the value of $\varepsilon$ in \eqref{eq:closed-form-power-updated} is determined by $\varepsilon=\rho_j\varepsilon_v$ where $v\in \{1,~2,~3\}$.

\begin{figure}[t]
\includegraphics[width=1.1\columnwidth,trim={3.08cm 0 0cm 0},clip]{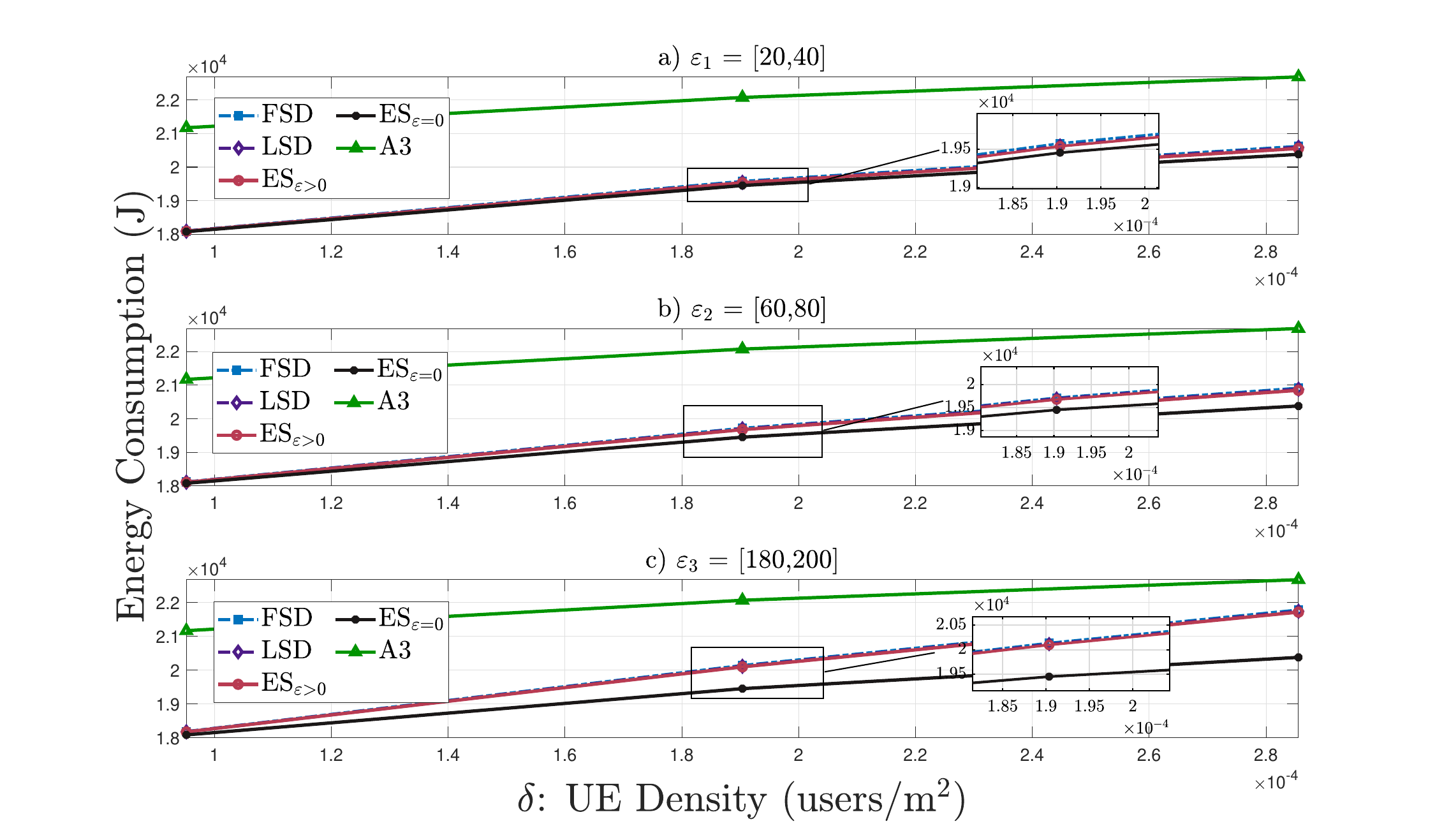}
    \caption{Energy consumption for $n=4$ SBSs with three different $\varepsilon$ values.}
    \label{fig:4BS}
\end{figure}

Fig.~\ref{fig:4BS} shows the energy consumption results with respect to $\delta$ for three different $\varepsilon$ values when $n=4$.
The first takeaway from these results is that as $\varepsilon$ increases, the differences between the values in ES$_{\varepsilon=0}$ and the other algorithms become higher.
This is because the error is unidirectional (i.e., it always adds up on the load: an assumption) and thus when it goes higher the energy consumption rises as well due to the relation in~\eqref{eq:closed-form-power-updated}.
Second, both FSD$_{\varepsilon>0}$ and LSD$_{\varepsilon>0}$ yield almost the same results with the ES$_{\varepsilon>0}$; there is a \textit{maximum of $0.3\%$ relative difference} between both $Q$-learning designs and ES$_{\varepsilon>0}$ for all $\varepsilon$ and $\delta$ values. 
As FSD$_{\varepsilon>0}$ and LSD$_{\varepsilon>0}$ give almost the same results, they seem overlapped in Fig. \ref{fig:4BS}.
This showcases how accurate our $Q$-learning designs work, considering that the ES algorithm is optimum.
Comparing the $Q$-learning designs with each other, the simulation results demonstrate that LSD$_{\varepsilon>0}$ gives its closest performance to FSD$_{\varepsilon>0}$ for $\varepsilon_1$ and $\delta=9.5\times 10^{-5}$ values, and the relative difference between LSD$_{\varepsilon>0}$ and FSD$_{\varepsilon>0}$ diverges up to $0.16\%$ as $\varepsilon$ and $\delta$ rise.
However, this is still a good result given that we carry the LSD algorithm with less computational cost and it manages to reach almost the same result as FSD.
When A3 and ES$_{\varepsilon=0}$ are compared, the maximum relative difference is obtained as $14.6\%$ for $\delta=9.5\times 10^{-5}$.
Moreover, the relative difference between A3 and ES$_{\varepsilon>0}$ is found to be $14.54\%$ for $\varepsilon_1$, which becomes $14.18\%$ when $\varepsilon_3$ takes place for the same $\delta$ value. 
As $\delta$ increases, the relative differences between A3 and both ES$_{\varepsilon=0}$ and ES$_{\varepsilon>0}$ cases keep decreasing down to $10.15\%$ and $4.24\%,$ respectively.
The inference behind this is that as the network becomes denser, all the CS algorithms behave closely with A3 because it is not wise to switch off an SBS due to the high demand.

\begin{figure}[t]
\includegraphics[width=1.1\columnwidth,trim={3.08cm 0 0cm 0},clip]{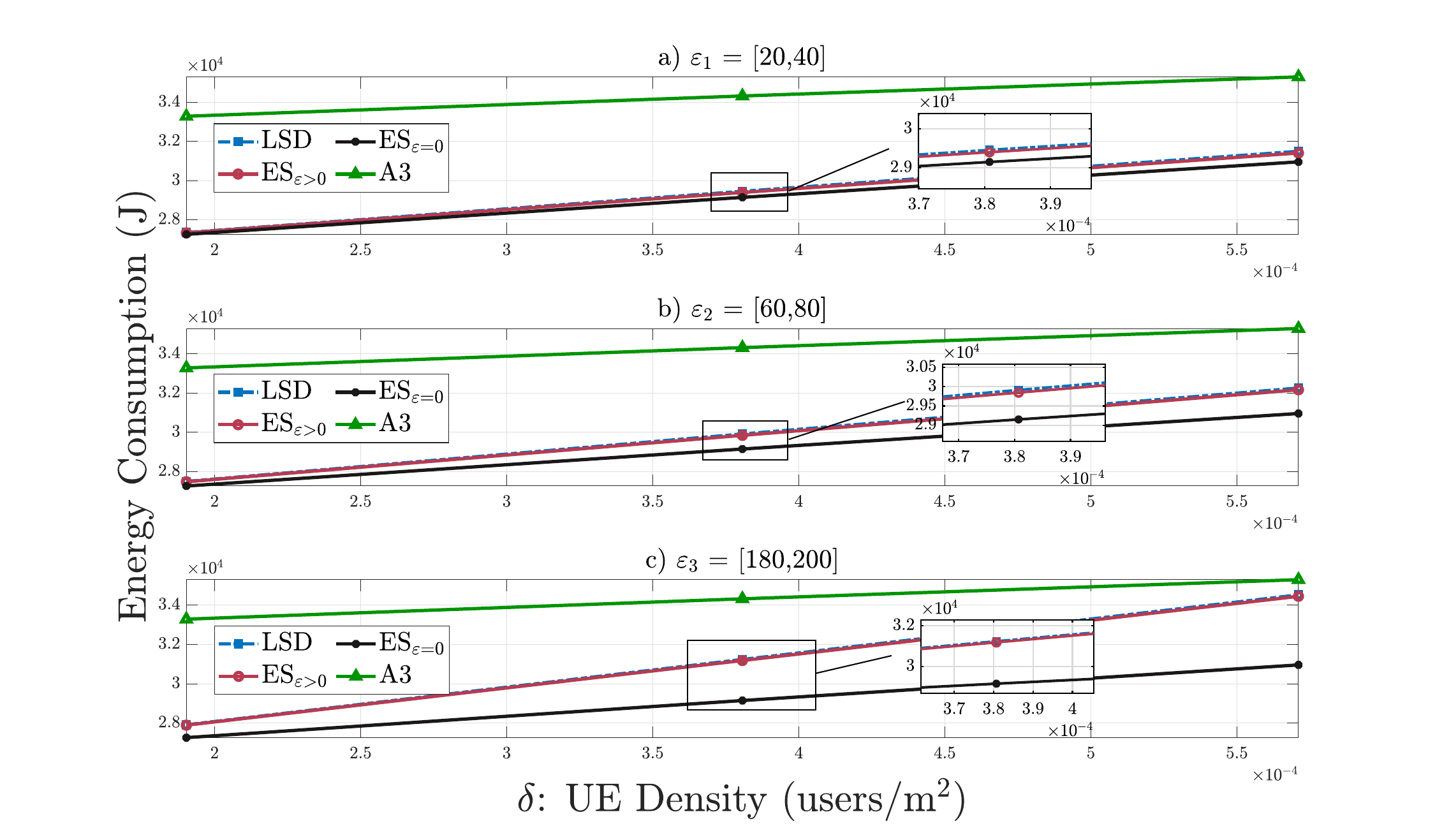}
    \caption{Energy consumption for $n=8$ SBSs with three different $\varepsilon$ values. }
    \label{fig:8BS}
\end{figure}

Fig.~\ref{fig:8BS} demonstrates the energy consumption results with respect to $\delta$ values when $n=8$.
Since the network is larger in size now, we double the number of UEs, therefore $\delta$ values are multiplied as well, as the size of the environment remains the same.
The results keep a similar trend with the outcomes in Fig.~\ref{fig:4BS}.
Another similarity is that LSD performs very close to ES$_{\varepsilon>0}$ when $n=8$.
The results illustrate that the maximum and minimum relative differences between LSD$_{\varepsilon>0}$ and ES$_{\varepsilon>0}$ are found as $3.0\%$ and $0.11\%$, respectively, among all $\varepsilon$ and $\delta$ values.
These confirm that the LSD algorithm works well even when the network is denser.
Note that there is no FSD algorithm included in this set of simulations due to the implementation challenges.
When A3 and ES$_{\varepsilon=0}$ are put into the consideration, the maximum relative difference is obtained as $18.09\%$ for the $\delta=19.03\times 10^{-5}$.
The relative difference between A3 and ES$_{\varepsilon>0}$ is found as $18.04\%$ in the $\varepsilon_1$ regime, which declines as low as $17.76\%$ when $\varepsilon_3$ is in place for the same $\delta$.
As $\delta$ increases, the relative differences between A3 and \st{for} both ES$_{\varepsilon>0}$ and ES$_{\varepsilon=0}$ drop to $12.28\%$ and $8.8\%$, respectively.

From Theorem \ref{th:est}, we understand that the estimation error, $\varepsilon$, has the potential to alter the CS policy, $\pmb \eta$. 
This change, in turn, influences network traffic performance and user data rates, as users may become associated with sub-optimal BSs when the CS policy differs from the optimal one, denoted as $\pmb \eta_\text{opt}$, leading to $\pmb \eta \neq \pmb \eta_\text{opt}$.
This deviation from the optimal policy can result in both under-connection, where fewer users are connected compared to the optimal policy, and over-connection, where more users are connected. 
Such variations have a direct impact on network traffic performance.
Furthermore, as discussed in~\cite{bizim_icc}, an increase in $\varepsilon$ leads to a reduction in user data rates. 
This reduction occurs because users may be connected to BSs that do not provide the expected SINR performance, as dictated by the optimal policy $\pmb \eta_\text{opt}$. 
As a result, users experience lower data rates, as they typically select BSs with the best SINR performance.

\begin{figure}[h!]
\centering
\includegraphics[width=.65\columnwidth,trim={.2cm .2cm 1.2cm .6cm},clip]{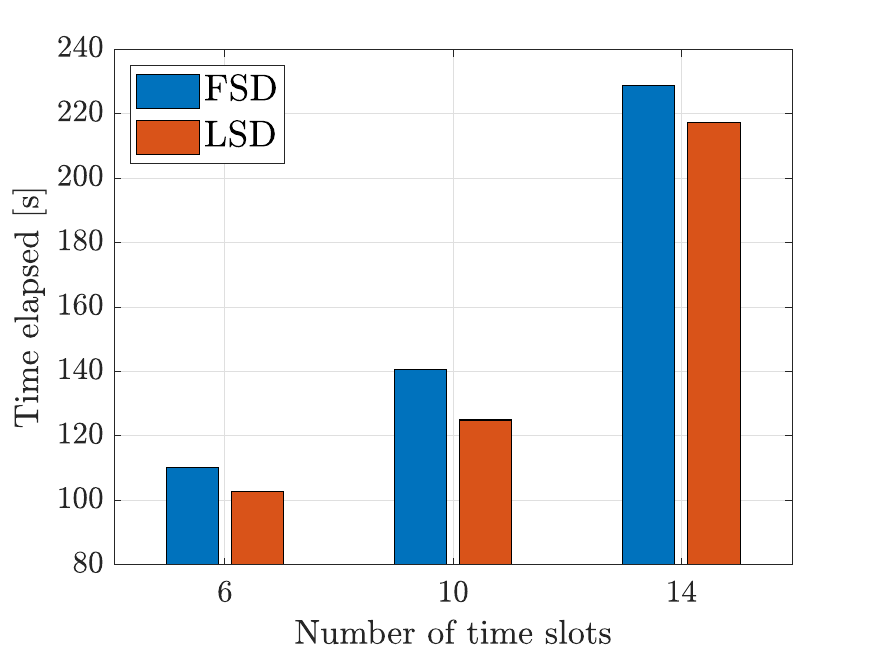}
    \caption{Elapsed time comparison between FSD and LSD for different number of time slots when $n=4$. Results are averages of 10 runs (Monte Carlo).}
    \label{fig:complexity}
\end{figure}
We conducted an empirical test on the complexities of both designs, as illustrated in Fig. \ref{fig:complexity}.
The simulations were performed using a computer with the following specifications: Processor: 12th Gen Intel(R) Core(TM) i5-1230U @ 1.00 GHz; RAM: 8GB (7.7 GB usable); System Type: 64-bit operating system, x64-based processor.
The results indicate that FSD results in higher complexity (i.e., elapsed time) compared to \st{the} LSD. 
However, the difference is not as significant as the theoretical computational complexity discussed in Section \ref{sec:proposed_q} using big-$O$ notation.
This observation is attributed to the fact that LSD and FSD differ primarily in the state design (i.e., dynamic part), while they share several common components such as actions and user-cell associations. 
As a result, the static part dominates the overall complexity, making the difference in the dynamic part less significant.
Furthermore, additional analyses reveal that the difference in complexity increases by approximately $21\%$ when the number of SBSs, $n$, increases from 4 to 8 due to the larger size of the $Q$-matrix.

\section{Conclusion}
The impacts of cell load estimation errors for CS operations in cellular networks were introduced for the first time in this study.
Moreover, two different $Q$-learning algorithms were developed, and in order to test the algorithms and observe the impacts of cell load estimation errors a VHetNet was considered, which includes the promising HAPS technology.
The results primarily confirmed the hypothesis behind the paper that the cell load estimation error is able to change the decision taken by optimization algorithms.
Further, both $Q$-learning algorithms performed reasonably well compared to the ES algorithm.
In future, we plan to develop solutions to estimate cell loads more accurately by means of artificial intelligence.


\bibliographystyle{IEEEtran}
\bibliography{output}
\end{document}